\documentclass{iau} 
\usepackage{graphicx}
\usepackage{natbib} 
\renewcommand{\cite}[2][1]{#1\nocite{#2}}

\title[Div.~E~~Synthetic activity indicators for M-type dwarf stars] 
{Synthetic activity indicators for M-type dwarf stars}

\author[Wedemeyer et al.]   
{Sven Wedemeyer$^1$
\and Hans-G\"unter Ludwig$^2$
}

\affiliation{$^1$Institute of Theoretical Astrophysics, University of Oslo,\\
 Postboks 1029 Blindern, N-0315 Oslo, Norway\\
 email: {\tt sven.wedemeyer@astro.uio.no}
 \\[\affilskip]
$^2$ZAH-Landessternwarte, University of Heidelberg , Heidelberg, Germany.\\
email: {\tt hludwig@lsw.uni-heidelberg.de}
}

\pubyear{2015}
\volume{320}  
\jname{Impacts of Solar and Stellar Flares on Atmospheres, Coronae and Winds}
\editors{A.C. Editor, B.D. Editor \& C.E. Editor, eds.}

\newcommand{\aj}{AJ}
\newcommand{\aap}{A\&A}
\newcommand{\apj}{ApJ}
\newcommand{\apjl}{ApJL}

\newcommand{\ssr}{Space~Sci.~Rev.}          

\begin{document}

\maketitle

\begin{abstract}
Here, we present a set of time-dependent 3D RMHD simulations of a M-dwarf star representative of AD Leo, which extend from the upper convection zone into the chromosphere. The 3D model atmospheres are characterized by a very dynamic and intermittent structure on small spatial and temporal scales and a wealth of physical processes, which by nature cannot be described by means of 1D static model atmospheres. Artificial observations of these models imply that a combination of complementary diagnostics such as Ca II lines and the continuum intensity from UV to millimeter wavelengths, probe various properties of the dynamics, thermal and magnetic structure of the photosphere and the chromosphere and thus provide measures of stellar activity, which can be compared to observations. The complicated magnetic field structure and its imprint in synthetic diagnostics may have important implications for the understanding and characterization of stellar activity and with it possibly for the evaluation of planetary habitability around active M-dwarf stars.

\end{abstract}

\firstsection 

\section{Introduction}
Cool red dwarf stars of spectral type M are the most abundant type of star in our galaxy       
and presumably in the whole universe (\cite[Bochanski et al. 2010]{2010AJ....139.2679B}). 
Studying these ``M-dwarfs'' in detail has therefore far-reaching implications for our understanding of stars in general. 
M-dwarfs exhibit flares that can be much stronger than their strongest solar analogues, exceeding the bolometric luminosity of the whole M-dwarf for minutes 
(see, e.g., \cite[Hawley \& Pettersen 1991, Kowalski et al. 2010, Osten et al. 2010, Schmidt et al. 2014]{Hawley1991,Osten2010,2010ApJ...714L..98K,2014ApJ...781L..24S}). 
Such energetic events require a large amount of energy, which would be stored in the magnetic field in the star's atmosphere and is then explosively released during a flare. 
Magnetic field strengths derived from observations imply that high (average) values of up to 3-4\,kG can be reached                
(see, e.g., \cite[Hallinan et al. 2008, Reiners \& Basri 2009, 2010]{2008ApJ...684..644H,2009ApJ...705.1416R,2010ApJ...710..924R}). 
Understanding  the formation of energetic flares would require knowledge about the magnetic topology of the flaring region on a star.  
Unfortunately, it is difficult to infer how the magnetic field is structured on the spatial scales relevant for flares based on spatially unresolved observations of stars. 
In contrast, high-resolution observations of our Sun have progressed our understanding of the magnetic topology in the solar atmospheres and the prerequisites of solar flares. 
It also facilitated the development of numerical simulation codes that can already reproduce many of the observed solar properties. 
The same codes, which have been tested on the solar reference case, can also be applied to other stellar types including M-dwarfs. 

\begin{figure}[th!]
\begin{center}
 \includegraphics[width=\textwidth]{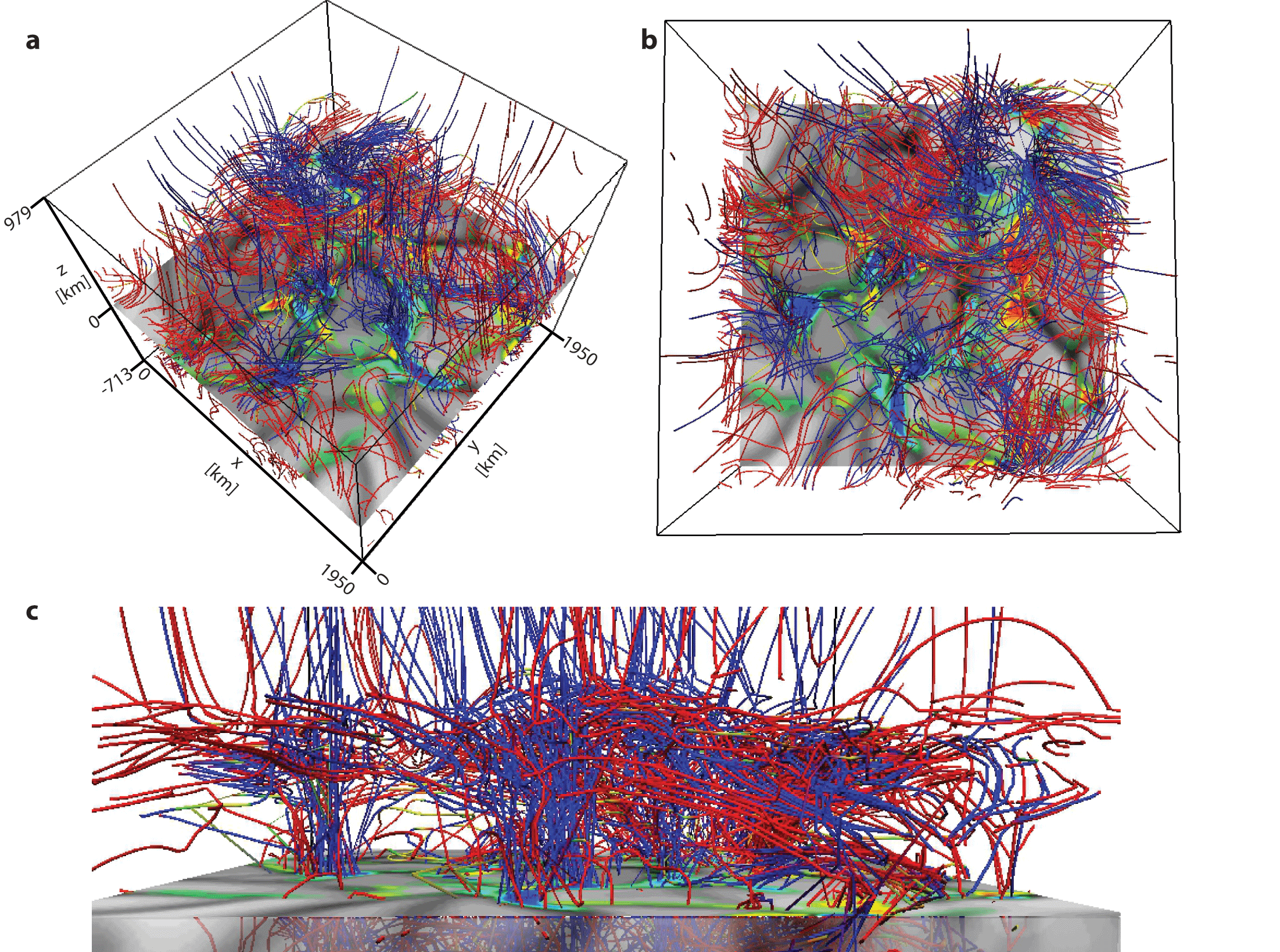} 
 \caption{Visualisation of a simulation snapshot from the mixed polarity model with $B_0 = 100$\,G  at \mbox{$t = 2550$\,s} after the magnetic field was inserted. 
The whole computational box is viewed from different angles in panels a and b, 
whereas a close-up from the side is displayed in panel~c. 
The vertical velocity at $\tau_\mathrm{R} = 1$ is plotted as grey shades, 
revealing the granulation pattern at the bottom of the photosphere. 
The same areas are overlaid with regions with high negative (red) and high 
positive (blue) values of the vertical magnetic field component $B_z$. 
The magnetic field lines, which are calculated from all spatial magnetic field 
components, connect patches of different magnetic polarity (red or blue) or reach 
the upper boundary as open fields. 
The field lines have colours corresponding to the local value of $B_z$, again with 
red for negative and blue for positive sign. 
The images are created with VAPOR (\cite[Clyne et al. 2007]{vapor_clyne2007}).}
   \label{wedemeyerfig1}
\end{center}
\end{figure}

Here, we present results from numerical simulations for a hypothetical M-dwarf which might be understood as an analogue of the well-studied star AD~Leo.  
Synthetic spectra, which are calculated based on these models, allow us to analyse how much information about the magnetic field topology is imprinted in spatially unresolved stellar spectra and which spectral indicators are best suited for probing the structure and activity of M-dwarf atmospheres.

\section{Numerical models}

The 3D radiation magnetohydrodynamic simulations presented here are calculated with CO5BOLD (\cite[Freytag et al. 2012]{2012JCoPh.231..919F}), 
a code which has been used for a wide range of stellar types, e.g., 
main sequence A-type stars (\cite[Steffen et al. 2005]{2005ESASP.560..985S}),  
AGB stars (\cite[Freytag \& H\"ofner 2008]{2008A&A...483..571F}), 
and even white dwarfs (\cite[Tremblay et al. 2015]{2015ApJ...812...19T}). 
Hydrodynamic M-dwarf models have been produced by 
\cite[Ludwig et al. (2002)]{2002A&A...395...99L}, \cite[Dorch \& Ludwig (2002)]{2002AN....323..402D} and \cite[Wende et al. (2009)]{2009A&A...508.1429W}, while \cite[Wedemeyer et al. (2013)]{2013AN....334..137W} produced magnetohydrodynamic M-dwarf models with chromospheres. 
The model presented here has an effective temperature of $T_\mathrm{eff} = 3240$\,K and 
a gravitational acceleration at the surface of $\log\,g = 4.5$, which is close to the values derived for AD Leo. 
The model extents from a height of -700\,km in the upper convection zone (i.e., from below the ``surface'')  to a height of +1000\,km in the chromosphere. 
The horizontal extent is 2000\,km\,$\times$\,2000\,km, which is large enough to fit several (photospheric) granules inside in each direction. 
A well-developed   initial hydrodynamic model with these dimensions was supplemented with an initial magnetic field with field strengths ranging from $|B_0| =10$\,G to 500\,G either in an unipolar vertical or in a mixed polarity setup. 
The magnetic field is quickly rearranged as the model atmosphere is advanced in time.

\section{Results}
\subsection{Structure and dynamics of M-dwarf models}

\begin{figure}[t!]
\begin{center}
\includegraphics[width=12.0cm]{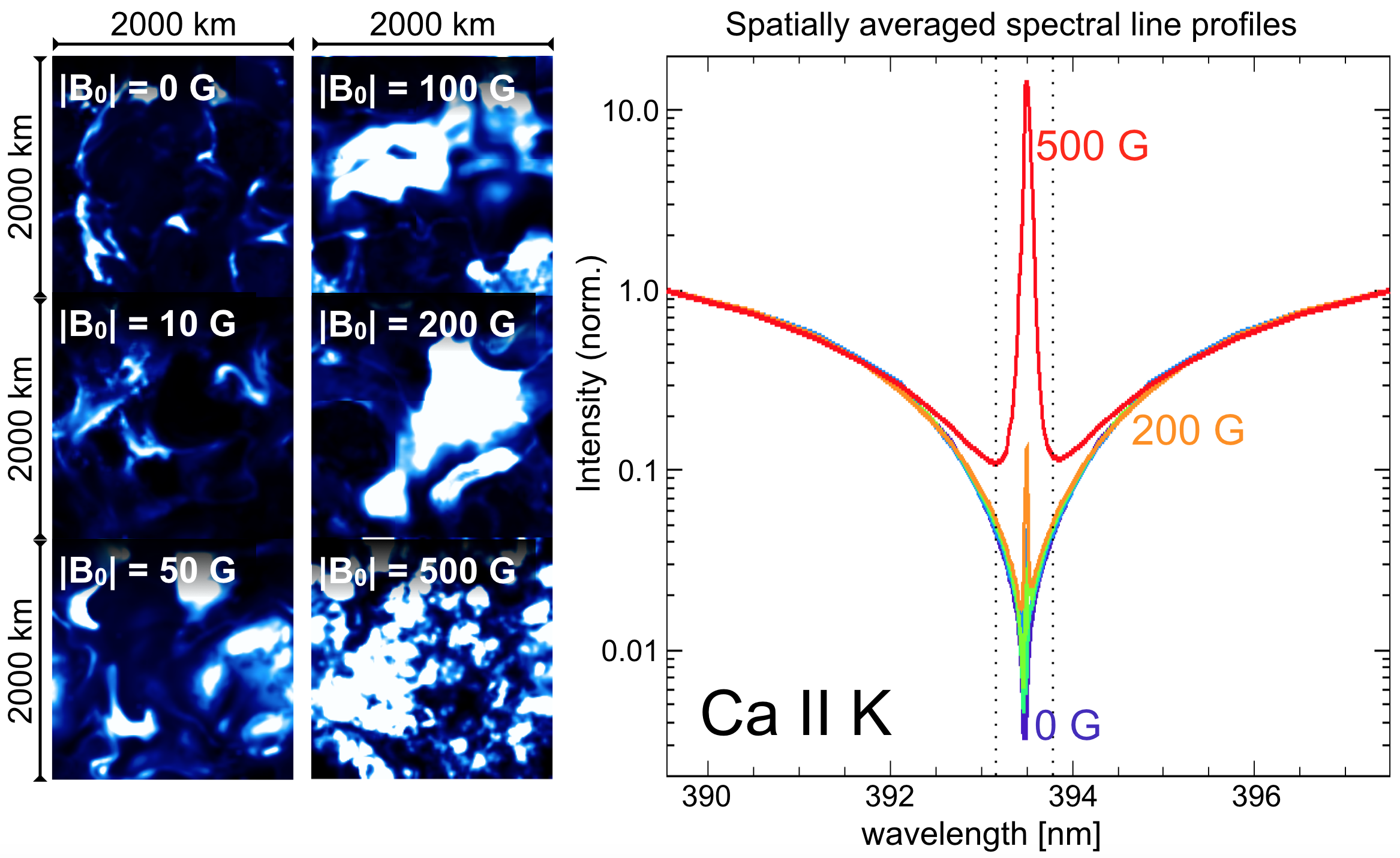} 
 \caption{\textit{Left:} Synthetic (disk-center) intensity maps in the line core of the Ca\,II\,K line for a selected time step of the models with different initial magnetic field strength $B_0$, representing a sequence with increasing magnetic activity level. 
\textit{Right:} Spatially averaged stellar spectra as function of wavelength for all 6 considered models. 
}
   \label{fig:wedemeyerfig2}
\end{center}
\end{figure}

In the photosphere, the magnetic field is essentially ``frozen-in'' and therefore rearranged by the convective flows. 
Consequently, the initial magnetic field is concentrated and in the intergranular lanes where it forms sheets and knots with magnetic field strengths with maximum values in excess of 1\,kG. 
This process takes place on short timescales related to the convective flow pattern and thus the lifetimes of granules, which is on the order of less than 5\,min. 
The models with lowest initial field strength $|B_0|$ produce only a few of these strong magnetic field concentrations in the photosphere whereas the models with high $|B_0|$ feature a high coverage of strong magnetic field concentrations. 
In the chromosphere above, the magnetic field lines funnel out and/or connect with photospheric magnetic footpoints of opposite polarity. 
The resulting complex magnetic field topology is visualized for the model with an initially mixed-polarity field with $|B_0| = \pm 100$\,G in Fig.~\ref{wedemeyerfig1}.   
Next to open magnetic field lines extending to the upper boundary of the model, small-scale magnetic loops form. 
The loop tops are located at the transition layer between photosphere and chromosphere as a result of convective overshooting, which produces regions with only weak magnetic field directly above the granules. 
This ``small-scale canopy'' is a phenomenon known from corresponding simulations of the Sun (\cite[Wedemeyer-B\"ohm, Lagg, \& Nordlund 2009]{2009SSRv..144..317W}) and coincides there with the ``classical'' temperature minimum region.

The M-dwarf models resemble solar models in many aspects but on smaller spatial scales and with smaller variations of atmospheric properties such as the gas temperature.  
The granules in the M-dwarf models have typical sizes on the order of $\sim 200$\,km as compared to $1000-1500$\,km in solar models. 
Photospheric vortex flows are found in both cases \cite[(Wedemeyer et al. 2013)]{2013AN....334..137W}.  
The temperature variations in the low photosphere are very small, which is also reflected in only small values for the contrast of the continuum intensity at visible wavelengths. 
The models exhibit propagating chromospheric shock waves but the peak temperatures reach only values on the order of 5500\,K, which is thus lower than the 7000-8000\,K typically seen in solar models.

\subsection{Synthetic observables}

\begin{figure}[t!]
\begin{center}
\includegraphics[width=12.0cm]{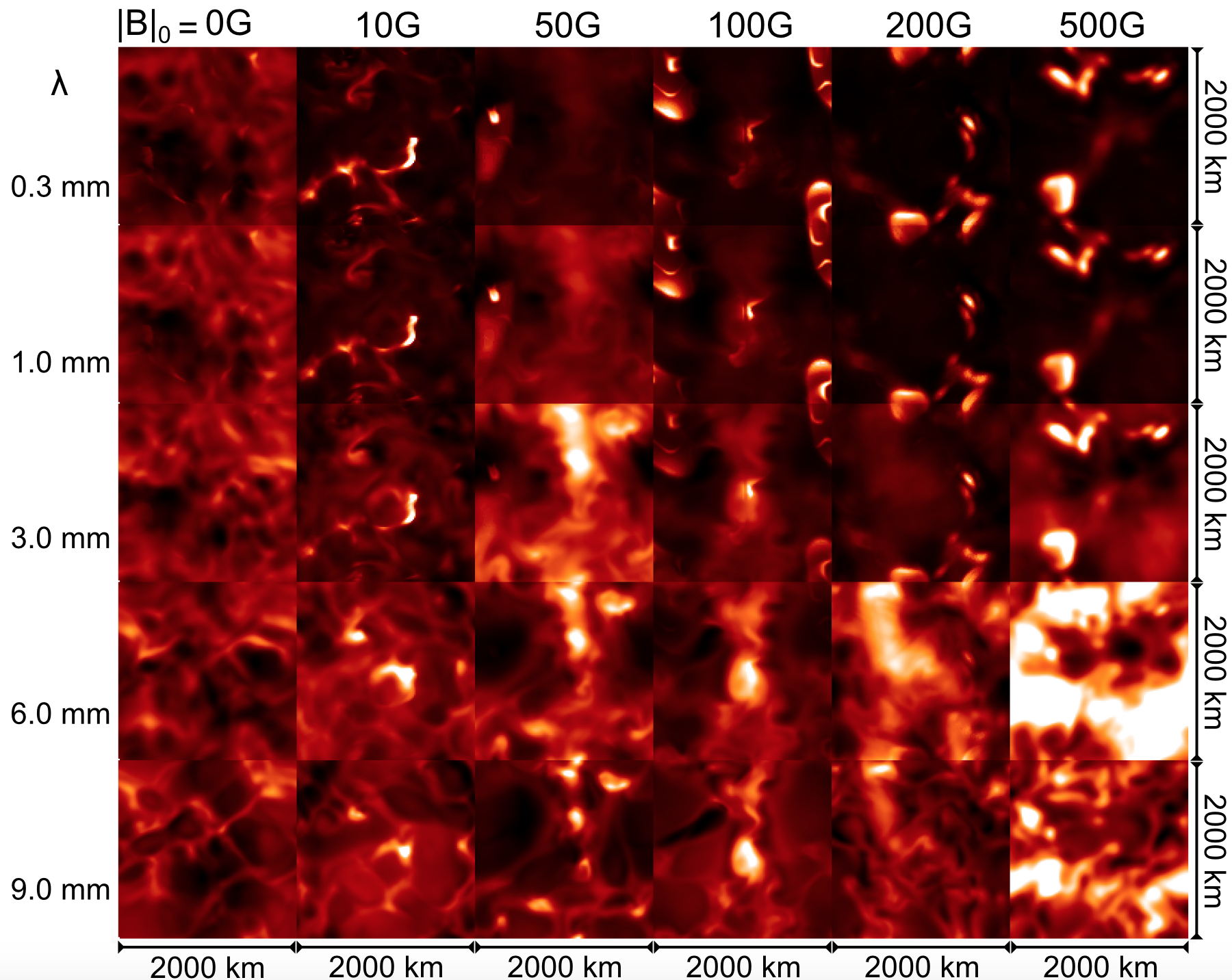} 
 \caption{Disk-center continuum intensity maps for wavelengths ranging from 0.3\,mm to 9.0\,mm (top to bottom) and for the increasing initial magnetic field strength $|B|_0$ and thus activity level of the model. 
}
   \label{fig:wedemeyerfig3}
\end{center}
\end{figure}

The aforementioned numerical models are used as input for radiative transfer codes, which produce corresponding synthetic observables such as continuum intensity maps and spectra. 
The example in Fig.~\ref{fig:wedemeyerfig2} shows the results of non-LTE calculations with MULTI (\cite[Carlsson 1986]{Carlsson1986}) for the Ca\,II\,K line. 
The maps on the left display the intensity in the line core as seen from above (i.e., at stellar disk center, $\mu = 1.0$) for a selected time step from each of the different models.  
The non-magnetic model ($|B_0| = 0$) exhibits only a few bright filamentary regions on top of a much darker background, which is caused by the interaction of hot, propagating shock waves in the stellar chromosphere.  
The area fraction of the bright regions increases with $|B_0|$ until a significant part of the computational box is filled with strong line core emission for the case with $|B_0| = 500$\,G. 
Accordingly, the horizontally averaged spectrum changes with $|B|_0$, i.e. with activity level of the local region as can be seen in the right part of Fig.~\ref{fig:wedemeyerfig2}. 
The central emission peak increases most notably with $|B|_0$ from being basically absent for the non-magnetic case to exceeding the continuum level for the model with 
$|B_0| = 500$\,G.

Stellar observations at wavelengths between 0.3\,mm and 9.0\,mm with the Atacama Large Millimeter/Submillimeter Array (ALMA, see, e.g., \cite[Wooten et al. 2009]{2009IEEEP..97.1463W}, \cite[Liseau et al. 2015]{2015A&A...573L...4L}) provide diagnostics complementary to Ca\,II\,K because the intensity at ALMA wavelengths probes different plasma properties. 
Continuum intensity maps at these wavelengths for the M-dwarf models are shown in Fig.~\ref{fig:wedemeyerfig3}. 
The maps are calculated with LINFOR3D (see \mbox{http://www.aip.de/$\sim$mst/linfor3D$\_$main.html)}, which solves the detailed radiative transfer in 3-D under 
the assumption of local thermodynamic equilibrium (LTE). 
The assumption of LTE is valid at millimeter wavelengths because the opacity is due to 
particle interactions and the source function of the radiation continuum is thus Planckian.
As for the Sun, the height range in the stellar atmosphere from where the (sub)-mm intensity emerges increases the wavelength (\cite[Wedemeyer-B\"ohm et al. 2007]{2007A&A...471..977W}). 
The intensity at the shortest wavelengths originates from the top of the photosphere/the low chromosphere whereas the intensity at the longest wavelengths originates from the upper chromosphere. 
Comparing disk-center maps at different wavelengths (columns in Fig.~\ref{fig:wedemeyerfig3}) thus gives important information on the stratification of the model atmospheres. 
As for the Ca\,II\,K line, the area fraction filled with bright emission increases with the $|B|_0$. 
However, in contrast to the Ca\,II\,K line, the continuum intensity at ALMA wavelengths is basically a linear measure of the gas temperature in the probed atmospheric layer and should thus be easier to interpret.


\section{Discussion and Outlook}

Already the first model generation clearly shows that M-dwarf chromospheres have a complicated,  intermittent and very dynamic 3-D structure, which can certainly not be described by one-dimensional static model atmospheres. 
Considering the intermittent nature of these chromospheres has therefore important implications for the interpretation of observations and our understanding of the atmospheres of cool stars. 
A central question is how much information about the small-scale atmospheric structure can be unambiguously extracted from spatially unresolved stellar observations. 
In view of the intricate structure we also have to carefully ask what an observed 
(spatially averaged) intensity tells us. 
The answers to these questions depend strongly on the employed diagnostic because the 
spatial average of an observable that depends non-linearly on the properties of the atmospheric gas (e.g., the gas temperature and electron density) is in general not the same as the observable corresponding to the spatially averaged atmospheric properties
(see, e.g., Fig.~3 in \cite[Wedemeyer et~al. 2013)]{2013AN....334..137W}. 
In that sense, the preliminary results presented here already demonstrate the potential of sets of complementary diagnostics that probe different properties of the atmospheric plasma, like, e.g.  Ca\,II\,K + H$\alpha$ + (sub-)mm continua.

Next to observations of the Sun as a star (see, e.g., \cite[Dumusque et al. 2015]{2015arXiv151102267D}),  detailed numerical simulations of stellar atmospheres are essential for understanding and developing the necessary diagnostic tools and strategies. 
The next steps will be (1)~to populate a stellar surface with regions of different activity level (i.e., models with different $|B_0|$), (2)~to calculate the resulting stellar spectrum and (3)~to try to extract as much information about the thermal and magnetic structure of as possible.  
Furthermore, the pronounced inhomogeneities of the model atmosphere make it necessary to take into account the 3D nature of the radiation field as it has been demonstrated by 
first successful calculations with PHOENIX3D (\cite[Hauschildt \& Baron 2010]{2010A&A...509A..36H}). 
More detailed publications about the model atmospheres and synthetic observables are currently under work.\\[3mm] 
\indent S.~W. was supported by the Research Council of Norway (grants 208011/F50 and
221767/F20). 
H.G.L. acknowledges financial support by the Sonderforschungsbereich SFB 881 \textit{``The Milky Way System''} (subproject A4) of the German Research Foundation (DFG).

\bibliographystyle{aa} 

\end{document}